# A guided map to the spiral arms in the galactic disk of the Milky Way

Jacques P. Vallée

National Research Council Canada, Herzberg Astronomy & Astrophysics
5071 West Saanich Road, Victoria, B.C., Canada V9E 2E7
email: jacques.vallee@nrc-cnrc.gc.ca   phone: 1-250-363-6952    fax: 1-250-363-0045
ORCID number   http://orcid.org/0000-0002-4833-4160

**Abstract.**

An up-to-date overview of the recent history is given, aiming to present a helpful working guide to the literature and at the same time introduce key systems and observational results, starting from the Sun and going towards the Galactic Center and parts of the *Zona Galactica Incognita*, and beyond.

We start by presenting an observational view of the Milky Way's disk plane (cartographic, dynamical, chemical cross-cut, magnetic). This included the four long spiral arms in the disk of the Milky Way galaxy (Fig.1), their geometry (Fig.2), components, velocity (Fig.3), their widths and internal layers as well as onion-like ordered offsets (Fig.4), the central galactic bars, arm tangents, arm pitch and arm shape, arm origins near the Galactic Center (Fig.5), and other possible players in the spiral arm, such as the magnetic field (Fig.6) and the dark matter content.

After, we present a basic analysis of some theoretical predictions from galactic arm formation: numerical simulations or analytical theories, and observations are checked against predictions from various numerical simulations and analytical (theoretical) models.

## 1. Introduction and scope – the Galactic disk domain

Our galaxy has perhaps a trillion stars, orbiting around their common center in an ordered way.  Knowing  better our Galaxy observationally is our aim, as well as making basic physical explanations, and making preliminary comparisons to some predictions from numerical simulations or analytical models. Aside from irregular galaxies, interacting galaxies, and dwarf galaxies, the remainder consists of elliptical galaxies (about 1/3 of all galaxies)  and spiral galaxies (about 2/3)  - see Dobbs & Baba (2014).  A good minority of spirals have a flocculent or asymmetric shape without long arms  (30%), while the remaining  spirals have a symmetric shape with either 2 long arms (10%) or else with 4 or 6  long arms  (60%) – see Elmegreen & Elmegreen (1982; 1989).  What about the Milky Way galaxy itself?  Most astronomers answer that the Milky Way is approximately symmetric spiral with 4 long arms.

Here we concentrate mainly on the gaseous and stellar disk of our spiral galaxy the Milky Way, while briefly sketching the galaxy's halo and  nucleus.  For a review of the integrated properties of the Milky Way, see Bland-Hawthorn and Gerhard (2016) although their Figure 11 with spiral arms dates back a decade ago to Momany et al (2006 – their fig.3).  For a grand review of the theories of spiral arm formation and evolution, the reader is referred to Dobbs and Baba (2014) and Shu (2016).

To get going, we start geometrically (Section 2), employing galactic coordinates and logarithmic spiral arm equations, along with a distance scale, to bring pitch angle and interarm separation, leading to a discussion of techniques to deal adequately with some observational issues (say, how to appraise the foreground and the background and remove them, to leave only the spiral arm under study).

We then embark on the basic velocities involved for these arm features (Section 3), employing an assumed velocity dispersion, orbital circulation, and arm position in the velocity-longitude domain, leading to a discussion of techniques to deal adequately with a few observational issues (say, how to appraise dust extinction, being different at different galactic longitudes). Having both the geometry and the velocity distribution, one can enter the chemical domain (Section 4), employing different arm tracers as seen with telescopes, along different galactic longitudes and velocities, bringing statistics to deal with small error bars in measurements, leading to a discussion of techniques to deal adequately with a few observational issues (say, whether to give more weight to some chemical arm tracers, instead of the same weight).

Having reviewed the geometry, velocity and chemistry within a 5-kpc radius around the Sun, one proceeds to the interior of the Galaxy, namely within a 3-kpc radius of the Galactic Center (Section 5), starting with observational data about the start of each spiral arm there, and bringing the so-called '3-kpc-arm' features in the fold, leading to a discussion of techniques to deal adequately with a few observational issues (say, the 'blurred' molecular ring seen by poor angular-resolution telescopes, regarded by some to be composed of separate molecular clouds as seen at specific galactic longitudes).

Some basic explanations for the spiral arms follow (Section 6), employing the predictions of numerical simulations and analytical theories, bringing in the angular rotations of galactic features (including the central bars), and comparing with observational data, leading to a discussion of techniques to deal with a few observational issues (say, the number of coexisting different density waves, if any). Comparing all theoretical predictions with all observations is outside the scope of this review.

We then look around for other, secondary features, that could perhaps be found later to have a bearing at some level on the spiral arms (Section 7), including a weak magnetic field in the galactic disk, or else a small quantity of dark matter in the spiral arms or interam, leading to a discussion of techniques to deal adequately with a few observational issues (say, the pitch angle of the stellar arm being the same or close to the pitch angle of the magnetic field in the arms).

## 2. Condensed survey of arm location, shape, pitch angle, interarm, distance

In this Section, the aim of this chapter is to provide the reader with an up-to-date global view of the Galactic disk domain, how to trace the arms in a logarithmic fashion around the Galactic Center using observable parameters, where are the interarms and the Sun, and how to determine the distance to a galactic object.

Current maps of the Milky Way disk can differ somewhat, much like 16$^{th}$ Century maps of the Earth's continents also differed, with unknown parts labelled "Terra Incognita". Thus was defined a "*Zona Galactica Incognita*" in the map of spiral arms for areas beyond the Galactic Center (Vallée 2002 - fig.2 and fig.3).

2.1 Global view.

The Galaxy has a central black hole, an elongated thick nuclear bulge bar and two other thin exotic galactic bars (a 0.5-kpc-radius nuclear bar – see Table 2 in Vallée 2016a; a 4.2-kpc-radius long bar – see Table 3 in Vallée 2016a) , and there is a huge disk around the galactic center (composed mainly of stars, dust, gas, mainly organized in spiral arms), and surrounded by a galactic halo. On a larger scale, the outer Milky Way seems to warp a little. Near l= 90$^o$ the mean disk is elevated to a galactic latitude b= +1$^o$ (Vallée, 2011 - fig. 38) while in the opposite direction l=270$^o$ the mean disk sinks to a latitude near b= -1$^o$.

Observations of the Milky Way galaxy are tricky, as the Sun is near the middle of the galactic disk, so a view from Earth shows all the arms along the same narrow band on the sky (most arms seen at Earth are behind another arm).

The spiral structure can be determined observationally. A large number of localized observational efforts were made to observe the Milky Way in various segments of galactic longitudes. Many pieces of sky were scrutinized, one piece at a time, and their data were extracted and published. For just one example, see Vallée (2002).

Rather than geometric space, some have employed phase space. Some have separated the stars in the disk into phase spaces (spatially, kinematically, chemically), or into a group of older stars with high metallicity and high above the mid-plane, and a group of stars with low metallicity closer to the mid-plane (Hawkins et al 2015).

Observers seek to obtain measurable parameters of spiral arms. One goes to a telescope operating at a wavelength (λ) or a spectral frequency (f), and do measurements of the intensity (I), polarization state (q) of a chemical tracer (t) in an object located at a given galactic longitude (l), latitude (b) and radial velocity (v). If the object is towards a spiral arm seen tangentially, the object's tracer (extended CO, say) would appear intense. In addition, foreground and background radiation would enter the measurements. Thus the measurements (I, f, q) of an object in arm Y (l,b,v) seen tangentially would also contain data from arm X in front along the same line of sight, as well as the interarm gas and the background disk emission.

2.1.1 Sun to Galactic Center distance

Early astronomers, up to and including Herschel in the late 1800s, thought that the Sun was at the centre of our Galaxy. It took Shapley in the early 20$^{th}$ Century to separate the sun's position from the Galactic Center's position, by about 20 kpc (Vallée 2005 - fig. 2). Modern observations shows that the Sun is separated from the Galactic Center by about 8.0 kpc, give or take 0.2 kpc (Genzel et al 2000; Table 3 in Bland-Hawthorn and Gerhard 2016). This allows modern theories to properly present the Sun near but outside a spiral arm in the galaxy.

**Table 1** shows very recent attempts at measuring the distance of the Sun to the Galactic Center [GC], and the mean circular rotation velocity of stars near the Sun as they orbit around the GC. Each column of Table 1 is described here: the first column refers to the distance between the Galactic Center and the Sun, in kiloparsecs. The 2$^{nd}$ column refers to the kinematic circular orbital velocity of objects near the Sun around the Galactic Center, excluding peculiar velocities. The 3$^{rd}$ column refers to the observational objects measured. The last column gives the appropriate references.

2.1.2 Spiral arm shape

Among nearby spiral galaxies and in the Milky Way, many observed arm shape are logarithmic. This is the spiral for which the radius R grows exponentially with the angle θ. Still, other shapes have been reported, such as an incomplete ring, an incomplete ellipse, a short line, or a complex polynomial.

The basic model of a Galaxy, with a spiral arm intensity I at galactic radius R and angle θ using a coordinate system based at the GC, with 'm' different arms, a logarithmic 'ln' shape, an arm pitch angle 'p', follows from the equation (from Vallée 1995, equ.1):

$$I(\theta, R) = A \cos \gamma \qquad (1)$$

$$\text{where } \gamma = [ \, 0.5m \, ( \, \theta - \theta_o - [(\tan p)^{-1} \ln(R / R_o) \, ] \, ) \, ] \qquad (2)$$

and $\theta_o$, A, and $R_o$ are constants to be fitted, $p > 0^o$ with $\theta > 0^o$ for an inward going arm (an arm spiraling toward a larger galactic radius 'R' as one proceeds to a greater angle θ, defined from the x-axis starting at the Galactic Center and going parallel to galactic longitude $90^o$).

Histograms for the 66 published pitch angle measurements before 2005 (Paper III) can be compared to histograms for 94 recent publications since 2015 (Paper X – Vallée 2017b), and a higher peak and a narrower base can be found in the recent histogram, showing that a convergence is happening. Ditto for the published interarm measurement s, number of spiral arms m, and spiral arm shape; there is a convergence in that twelve-year time period (central peak being about 3 times higher now, base width being about 3 times narrower now).

**Figure 1** shows a sketch of a four-fold-symmetry spiral arm model adopted by the author in a series of meta-studies, and this model is in approximate agreement with most recent models.
This figure gives a rough conceptual knowledge of where stellar arms are located in the Milky Way disk, separately for the outer galaxy (fig. 1a) and for the inner galaxy (fig.1b).

Inputs are: m =4 arms, pitch angle p = $13^o$ inward with respect to a circle around the Galactic Center, logarithmic shape, and a distance of the sun to the Galactic Center $R_{sun}$ of 8.0 kpc. Also drawn is the short thick galactic bar, although the details are not yet agreed upon (2.1-kpc radius, at $25^o$ from the line of sight – see table 3 in Vallée 2016a). The interarm distance 's', measured vertically across the sun from the Sagittarius to the Perseus arm, is about 3 kpc.

For the spiral arms, a series of blocks, each with 15 to 20 publications, were collated and statistically analyzed. Then one needs a meta-analysis to mosaic them and to reconstruct a full view or model as seen from above – see Vallée (1995 – Paper I), Vallée (2002 – Paper II), Vallée (2005 – Paper III), Vallée (2008a – Paper IV), Vallée (2013a – Paper V), Vallée (2014a – Paper VI), Vallée (2014b – Paper VII), Vallée (2015 – Paper VIII), Vallée (2016a – Paper IX), Vallée (2017b – Paper X).

2.2 Arm Pitch angle
Thus the intensity 'I' is maximum when $\gamma = 0$, and the arm separation 's' between two adjacent arms (s = $R_2 - R_1$) at a fixed angle $\theta_1$ is (from Vallée, 1995 – equ. 7):

$$2\pi \, \tan p = m \, \ln(1 + [s/R_1]) \qquad (3)$$

The equations above define the Norma arm with $\theta_o = 0^o$, say. Owing to an imposed similarity conditions, each arm is offset from the preceding one by $90^o$ in galactic longitude, so for the Perseus arm $\theta_o = -270° + \theta_{o\,(Norma)}$, while $\theta_o = -180° + \theta_{o\,(Norma)}$ for the Sagittarius arm, and for the Scutum arm $\theta_o = -90° + \theta_{o\,(Norma)}$.

The use of equation (3) with these values of m, s, and $r_I$ (as the Sagittarius-GC distance) permits p to be obtained. Inside the Sun's orbit, one can draw from the Sun a tangent to a spiral arm.

Some arms can even show two arm tangents, one in Quadrant IV ($l_{IV}$) and one in Quadrant I ($l_I$), such as for the Carina-Sagittarius arm or for the Crux-Centaurus-Scutum arm, or for the Norma arm. It can be shown (Vallée 2015 - his equations 1 to 10 and his fig.1; Drimmel 2000 - his equation 1) that the following equation holds:

$$\ln[\sin(l_I) / \sin(2\pi - l_{IV})] = [\tan(p)] \; [l_I - l_{IV} + \pi] \quad (4)$$

This last equation yields p, when the two arm tangents to the same arm (in Quadrant I and Quadrant IV) are known. The last parenthesis at right in Equ. 4 is expressed in angles ($\pi$ is a half circle), all in radians. It is only a function of the observed quantities $l_I$ and $l_{IV}$. This method employs observed data over a wide range in galactic longitudes, and can be used separately for each arm tracer appearing in both quadrants (CO, dust, HII, etc).

Thus using equatin (4), for the twin-tangents to the Carina (longitude l near 284°) – Sagittarius (l near 50°) arm, the mean pitch angle p is 14.0° ±0.4° inward, while for the twin-tangents to the Crux-Centaurus (l near 310°) – Scutum (l near 31°) arm, the mean pitch angle p is 13.3° ±0.5° inward (see Vallée 2015, Tables 1 and 2). Moving to the twin-tangents to the Norma arm (l near 329°) and 'start of Norma' arm (l near 20°), the mean pitch p is 13.7° ±1.4° inward (Vallée 2017d), using the data in Vallée (2016b – table 3).

Because of the moderate pitch angle, no arm can turn around the Galactic Center more than one half of a full circle before reaching the Sun's orbit. A very small pitch angle would lead to very many turns of an arm around the Galactic Center, before reaching the Sun's orbit (Vallée 2014b, fig.4). Sitting near the Sun, looking towards the GC (galactic quadrants IV and I), arms with a very small pitch angle will develop many turns around the GC before reaching the Sun, and thus will display many arm tangents, yet most predicted arm tangents are never observed, and the other predicted ones do not match the observed tangents [table 3 in Vallée 2014b].

Some authors have expressed the view that a spiral pattern with a variable pitch angle is possible (Reid et al. 2014; Hou & Han 2014).

### 2.3 The interarm s

The Sun is located in an interarm, between the Sagittarius arm and the Perseus arm. In the more recent period from 2002 to 2016, the observed 's' value, calibrated with $R_{sun}$ near 8.0 kpc, turns out to be near 3.1 kpc (Vallée 2013a - fig.1; Vallée 2015).

### 2.4 Armlet, spur, bridge, branch

Near the Sun, there is a local 'armlet' (or 'spur' or 'branch' or 'bridge'). Many OB stars are found between l=60° and l=90°, within 2.2 kpc of the Sun (Uyaniker et al 2001). There may be such other spurs elsewhere in the Milky Way, as there are some similar armlets seen in other nearby spiral galaxies. Olano (2016) used the old (600 Myrs) stellar groups close to the Sun (their position and velocity) to explain the local 'armlet' through a model of an old supercloud (with a size near 600 pc) in orbit around the Milky Way, whose center collided with the Perseus spiral arm some 100 M years ago (his Figure 5), thus braking its speed (his Fig. 6) and curving its orbital path (his Fig. 12), now reaching the Sun's area in the interarm, while disintegrating over time to form the younger Gould's Belt and the local Orion armlet.

Near the Sun, different authors have drawn the local armlet at different orientation, position and length (see Bobylev and Bashkova 2014; Hou & Han 2014). A completely new interpretation of the Orion armlet is given by Lépine et al (2017).

Elsewhere in the Milky Way disk, one would expect some little armlets or branches aligned roughly parallel to the arms (at a small pitch), or else some little spurs or feathers aligned roughly perpendicular to the arms (at a large pitch).

Another interarm 'spur' (or 'armlet' or 'bridge') was found by Rigby et al (2016, their fig. 6 and sect. 3.4), covering $7^o$ in galactic longitude ($l=32^o$ to $39^o$) and 30 km/s in radial velocity (from $v_{lsr}$ = +60 to +90 km/s), located in between the Scutum arm and the Sagittarius arm.

Some small armlets are found alongside large arms in nearby spiral galaxies, unrelated to any grand design spiral.

2.5 Distance.

A patch of a spiral arm containing an object (a star, say), can be measured from the Sun (galactic longitude l, distance r), or else also from the Galactic Center ($\theta$, R).

**Figure 2** shows the Galactic quadrants (I to IV) around the Sun (small circle), with galactic longitudes $l=0^o$ to $360^o$ (shown as thick dots in a circle).

This figure gives the galactic longitude circle, galactic quadrant (I to IV) and radius r, as tied to the galactic coordinate system attached to the Galactic Center (x-y axis, radius R).

A spiral arm is shown, in a logarithmic shape, starting near the Galactic Center (square). An object (black star) on the spiral arm in Galactic quadrant II can be seen at longitude l and distance r from the Sun, with a radial velocity v (positive outward). Inside the sun's orbit, a tangent to a spiral arm ($l_{IV}$) has been drawn.

Distance determination is always done using some prior assumptions. Thus a kinematical model can be used to convert an observed radial velocity into a distance, assuming that the full measurement can be due to orbital motion around the Galactic Center. If a part of the radial velocity comes from another source, such as a supernova shell expansion or filament rotation, then the distance determination could be off. Different kinematical models exist, each using its own value of the circular orbital velocity at the solar orbit (between 200 and 250 km/s), thus adding a distance error of about 10% for a gas velocity of 225 km/s, say. Optical (UBVI) photometric calibration is reasonably known, using standard stars, color coefficients, zero-points data, DAOPHOT software; typical errors in distance can be 10%.

2.6 Summary

Several different methods have been devised to analyze the observations of a cloud, to perform a foreground and background subtraction, and to characterize only the data pertinent to that cloud. The background/foreground disk emission is fairly known ; the different CO models gave slightly different output values from roughly the same observed data, especially when using different bin sizes in galactic longitudes, or different model data cutoff (63% removal, the amount of disk emission with galactic longitude – see Fig. 14 in García et al 2014), or different complex disk emission model (Section 2 in Bronfman 1992).

Some prefer to create a 3-dimensional symmetric model for the emission from the embedded disk gas, and to fit its parameters to the observed baseline intensity in a function of galactic longitude, then subtract this disk emission from the observed data, leaving the residual as the arm gas (Sect. 2 in Bronfman 1992). Others use additional data such as the radial velocity to create a longitude-velocity plot, and to circumscribe the cloud there (within such permitted extent in velocity and permitted extent in longitude), then use that information to best fit that cloud in the intensity vs longitude vs distance diagram (García et al 2014, their fig. 9). Finally, a simple subtraction method is to use a second-order polynomial for the background / foreground disk emission on the plot of intensity versus galactic longitude (which is subtracted), and followed by using multiple Gaussians for the lines in the residual (Hou & Han 2015, their Fig.3); this method assumed universal dust behavior (neglecting the arbitrariness in the amount of mid-infrared dust extinction as a function of different galactic longitude – see Gao et al 2009, their Fig.6). Each method comes with a different number of assumptions (disk shape, velocity model, polynomial baseline, dust absorption, etc ), and with a different number of model variables to fit; the sum of the assumptions and model variables increases the final error bar.

### 3. Brief survey of arm kinematics and velocity

In this section, we provide the reader with a mid-scale (orbital) and a small-scale (turbulent) view of gas motions.

3.1 Small scale. Non-circular, localized gas motions can exist of the order of 10 km/s. These motions include interstellar shocks (from supernovae or encounters with a nonlinear density wave), expansion of superbubble shells, non-circular (elliptical) gas orbits around the Galactic Center, localized streaming motions from spiral density waves, infall of HI neutral gas from the halo, etc (Foster and MacWilliams 2006; Russeil et al 2007).

3.2 Mid scale. The stars and gas follow an orbit around the Galactic Center, obeying the gravitational attraction from the center of the Galaxy – much like the Earth and planets orbit the Sun. The gas and Sun will enter the slower moving arm (inner side), then exits the arm (outer side) some time later. Specifically, one assumes here that the Sun is interior to the corotation radius – a likely assumption.

Observations show a flat circular velocity (same at all radial distance from the Galactic Center). Adopting a flat circular velocity of 220 to 230 km/s above a minimum galactic radius of 2 kpc, one gets the basic equation for the radial velocity $v_r$ of an object at galactic radius r and galactic longitude l, as seen from the Sun (Englmaier and Gerhard 1999; Russeil et al 2007):

$$v_r = (\omega - \omega_o) R_{sun} \sin l \qquad (5)$$

with $(\omega - \omega_o)$ being the difference in angular rotation rate at r and at the Sun.

**Figure 3** shows the radial velocity as a function of galactic longitude. The 4-arm model in Figure 1 has a counterpart in this velocity-longitude diagram. Thus the radial velocity derived from the model of Figure 1 are shown in Figure 3.

For a chosen portion of an arm (colored), this figure shows the radial velocity (vertical axis) of the gas in that arm as a function of the galactic longitude (horizontal axis) of the arm portion.

Thus with a spectroscopic instrument mounted on a telescope on Earth pointed at a specific galactic longitude (in Fig.2), a gas or star's observed radial velocity can be measured, and one can infer its specific spiral arm (in Fig.3), and a distance from the Sun can be inferred for that arm at that longitude (in Fig.1).

3.3 Comparisons with the observations. The 4-arm model shows good agreement, in each Galactic Quadrant (GQ). Some of the comparisons below fall in a narrow range from the Galactic Center, and thus may not support any model employing circular orbits and giving near-zero radial velocities.

GQ I. Pandian & Goldsmith (2007 – their Fig.3) displayed the velocity of methanol masers in GQ I, and there is a good match in the 4-arm model (Vallée 2008a - Fig.3a). Stark & Lee (2006, their fig. 1) displayed the major $^{13}CO$ J=1-0 clouds for the Sagittarius arm (their C arm), the Scutum arm (their D arm), the Cygnus arm (their F arm), the Perseus arm (near l=30$^o$ - their A arm; near l=80$^o$ – their B arm), for which there is a good match to the model (Vallée 2008a - fig.3a). Zucker et al (2015) also found a good fit to this model (Vallée 2008a - fig. 3a) with the $^{12}CO$ data J=1-0 in the longitude ranges l=0$^o$ - 8$^o$ (their fig.29), l=8$^o$-16$^o$ (their fig. 26), l=16$^o$-24$^o$ (their fig.23), l=24$^o$-30$^o$ (their fig.19).

GQ II. Stark & Lee (2006 – their fig. 1) displayed the major $^{13}CO$ J=1-0 clouds for the Cygnus arm (their F arm), being a good match to the 4-arm model (Vallée 2008a - fig.3b). Dobbs et al (2006 – fig. 15b) displayed the observed $^{12}CO$ J=1-0 clouds for the Perseus arm and the Cygnus arm, for which there is a good match to this 4-arm model (Vallée 2008a - fig 3b).

GQ III. Yang et al (2002, their fig.5) displayed $^{12}CO$ J=1-0 gas for the Perseus and Cygnus arm, for which there is a good match to this model in Vallée (2008a - fig.3bc).

GQ IV. Brown et al (2014, their Fig.5) compared the HI absorption and radio recombination line velocities from 252 observed HII regions in GQ IV, and found most of them aligned with the 4-arm model. Kothes & Dougherty (2007) shows the HI profile in GQ IV, and their Fig.1 does compare well with this model for the line of sight at l=340$^o$, with similar intersections with the spiral arm encountered (Vallée 2008a - Fig. 3d). Also, their highest velocity HI gas (their fig.3) is at this model velocity of -140 km/s. Finally, they observed that the HI gas peaks between -140 and -100 km/s due to a "pile-up in velocity space", with gas distributed over a large distance interval due to seeing several arms in that longitude, as found in this model. Zucker et al (2015) also found a good fit to this model (Vallée 2008a - fig. 3d) with the $^{12}CO$ data J=1-0 in the Scutum arm, in the longitude ranges l=330$^o$-338$^o$ (their fig.37), l=353$^o$-360$^o$ (their fig. 33).

3.4 Discussion

There was an *issue* on the physical reality of the 'fingers' pointing at the Sun. In a supernova explosion, HI gas affected by the supernova will have a large front component and a large back component in radial velocity, which, if taken as orbital motion in a kinematic distance model, would be erroneously translated as HI gas at different distances along the line of sight, resulting in a 'finger' pointing at the Sun. At radio wavelengths, HI gas is measured in galactic longitude and radial velocity, and if a simple orbital velocity model is used to get its 'kinematical' distance, it will result in 'features elongated along the line of sight', like 'fingers' pointed at the Sun (HI in Levine et al 2006, their fig. 4a). Others studied more complex velocity models to get a more realistic distance from the Sun, using velocity deviations or jumps. Without these jumps, they reconstructed numerically some 'density fingers' pointing to the Sun (Baba et al 2009, their fig.10; Pohl et al 2008, their fig. 4).

At optical and near-infrared wavelengths, isochrone fits in the UBVRI method and variable reddening and absorption laws should differ for each galactic longitude (Fig. 6 in Gao et al 2009, between 3 and 8μm; Fig. 4 in Li et al 2015), due to variable dust density and dust size with increasing and decreasing galactic longitude, correlated with spiral arms. Using instead 'universal' dust laws (invariant with galactic longitude), then stars at the same distance, but at slightly different longitudes, could appear to go up or down in distance, along their specifically reddened line of sight at different galactic longitudes. They are seen as pseudo 'chains of O-B starts elongated along the line of sight' by Bobylev & Bajkova (2015b), visible in their Fig.4 at galactic longitudes $78^o$, $174^o$, $189^o$, and $290^o$, and in their fig.6 at longitudes $134^o$, $189^o$, $290^o$, $342^o$. These are seen as a pseudo line of open star clusters, near l= $242^o$ and near l=$258^o$ (Vázquez et al 2010, their fig.15; Vázquez et al 2008, their Fig.4). Other spiral galaxies have been observed, and spiral galaxies have never shown such 'fingers' pointing at a specific star in a specific location inside a galaxy.

Are there velocity jumps not due to distance offsets? Velusamy et al (2015) used the kinematical method, with the circular velocity of 220 km/s, to obtain the linear distances (their Table 1c) and linear offsets relative to the mid-arm (their Table 2a). They obtained maps in [CII] at λ158μm with Herschel HIFI as restored to an effective beam of 80" and a resolution of 2 km/s. They first looked toward the tangents to the spiral arm of Norma (near $328^o$ for the CO peak intensity) and toward the beginning of the Perseus arm (near $337^o$ for the CO peak intensity). These longitudes represent the cold CO (1-0) peak intensity lanes (Table 1 in Vallée 2014c). Thus towards the cold CO arm tangents, they found two [CII] peaks in radial velocity, separated by about 10 km/s (their fig.7b and fig.8b). They attributed their two [CII] peaks to one from a warm medium (compressed), and one from a colder medium (molecular). They identified the compressed medium towards the mid-arm as the "inner edge" of the spiral arm (their fig.11), although the observed dust lanes are farther away in longitudes, at an "innermost arm edge" (Table 1 in Vallée 2014c).

**4. The arm width, composition, structures in parallel lanes (physical offsets)**

When moving a telescope in galactic longitudes with a filter to observe a single tracer at a time (CO, dust, masers, etc), one canfind a peak intensity at a longitude where the arm is seen tangentially. The reader could then have a clear view of the onion-like separation of these arm tracers in galactic longitudes, as well as the reversal of these onion-like tracers as one crosses the Galactic Meridian (located at a galactic longitude of zero).

The aim here is to provide the observed mean offset in linear width of a tracer t with respect to tracer $^{12}$CO in a typical spiral arm, as well as individual offsets s in each individual arm. Few numerical or analytical theoretical predictions for the Milky Way conform to this offset reversal across the Galactic Meridian.

We can use different tracers to get the galactic longitude where an arm is seen tangentially.

4.1 Arm tangents

There is a recent published Catalogue of tangents of arm tracers (arms seen tangentially, using different chemical tracers) culled from the literature, and various statistics can ensue (Table 3 in Vallée 2016b). There are many wavelengths represented in this Catalogue of tangents to the spiral arms, as seen in gamma-ray, optical, near and far infrared, submillimeter, and radio. The Catalogue has a master table of the mean longitude for each arm and for each tracer (table 3), followed by individual tables (one table for each arm tracer) listing individual observational data for each arm (Tables 4 to 10 in Vallée 2016b). In its first publication, this Catalog had 43 entries (Table 3 in Vallée 2014a). In the most recent publication, it grew to 215 entries (Tables 3 - 10 in Vallée 2016b).

Uncertainties. For the $^{12}$CO tangent to the Crux-Centaurus arm, the mean of 309.5$^o$ has an r.m.s. of 1.0$^o$ and s.d.m. of 0.3$^o$ (Table 5 in Vallée 2016b). For the dust-870 -microns tangent to the Carina-Centaurus arm, the mean of 311.4$^o$ has an r.m.s. of 0.5$^o$ and a s.d.m. of 0.35$^o$ [table 9 in Vallée 2016b]. The separation ($^{12}$CO versus dust) is 1.9$^o$ and the combined s.d.m. is 0.46$^o$. The dust tangent is always closest to the GC than the CO tangent. For the $^{12}$CO tangent to the Sagittarius arm, the mean of 50.5$^o$ has an r.m.s. of 0.9$^o$ and s.d.m. of 0.5$^o$ (Table 5 in Vallée 2016b). For the dust-870 -microns tangent to the Sagittarius arm, the mean of 49.1$^o$ has an r.m.s. of 0.2$^o$ and a s.d.m. of 0.1$^o$ [table 9 in Vallée 2016b]. The separation ($^{12}$CO versus dust) is 1.4$^o$ and the combined s.d.m. is 0.5$^o$. The dust tangent is always closest to the GC than the CO tangent. Combining all the arms together, the angular separation ($^{12}$CO versus dust) is 3.2$^o$ with an r.m.s. of 0.68$^o$ and a s.d.m. of 0.3$^o$ (Table 1 in Vallée 2016b) giving a signal/noise ratio of 11, while the linear separation ($^{12}$CO versus dust) is 315 pc with an r.m.s. of 64 pc and a s.d.m. of 26 pc (Table 1 in Vallée 2016b) giving a signal/noise ratio of 12.

Using the dust tracer, the galactic longitude of each arm is always closer to the direction of the GC than for the CO tracer for that arm. This is true for ALL spiral arms [table 1 in Vallée 2016b]. Using the galactic quadrants, the hot dust is at a higher galactic longitude than the CO *before* (at left of) the Sun-Galactic Center line (in galactic quadrant IV), but the hot dust is at a smaller galactic longitude than the CO *after* (at right of) the Sun-Galactic Center line (in galactic quadrant I).

This 'reversal in ordering' across the Sun-GC line is predicted by the shocked physics of density waves. Thus the angular distance of a tracer (dust, say) from the arm center (CO) is positive towards the arm's inner edge (towards the Galactic Center), and negative otherwise (Vallée 2014c, note 2 to table 1; Vallée 2014a, note b to table 3). This observed reversal, around the Sun-GC line, is the first concrete, necessary and sufficient proof for the basic galactic density wave model with shocks.

It is important to use statistics in order to get the best galactic longitude of an arm tangent, from a given tracer, hence the presence of separate tables with their means and errors. Basic statistics with median and means, weights and error bars, balance out positive versus negative biases. Biases can arise notably from different model assumptions, differing procedures, differing averaging bin sizes, or different radial velocity range, or a smaller range in galactic longitudes (affected by the local emission from an object such as a supernova remnant), or different input data cutoffs.

To avoid overlapping when comparing different arms, it is important to 'anchor' all tracers to a common system; so one defines one of them (broad emission from $^{12}$CO 1-0) as the 'zero' of linear separation from the mid-arm. There is enough measurements in the CO catalogue (Vallée 2016b) to pinpoint accurately where the peak CO is located.

The motivation for choosing the galactic longitude of the CO peak intensity (observed at a low-angular resolution of 8' ) as the center of a spiral arm was explained (Vallée 2014a) thus: CO is prominent in intensity and well mixed with cold molecules, so a CO telescope with a broad angular view can easily find the tangent to a spiral arm having molecules, with a very small uncertainty in galactic longitude. This CO tracer is the low-excitation J=1-0 , low temperature (~10K), low-angular resolution (~8 arc-min), narrow-line width (~2 km/s), 115 GHz data from the Columbia survey, as integrated over a radial velocity range associated with a spiral arm (Grabelsky et al 1987; Bronfman et al 1988; Alvarez et al 1990; etc).

In our Galaxy, another good arm tracer is $^{13}$CO J=1-0. There is an absence of bright massive molecular clouds in the interarms (not observed in $^{13}$CO 1-0), as evidenced in Figures 12 and 13 in Roman-Duval et al (2009). The confinement of massive, bright molecular clouds to spiral arms implies that these clouds are formed in the arms, and have short lives (less than 10 Myrs), to prevent them from spreading in the interarms. Roman-Duval et al (2016) defined a 'diffuse' CO component as CO gas detected in the $^{12}$CO J=1-0 line but showing no emission in the $^{13}$CO J=12-0 line (their Sect.1, Fig. 9 and Table 5c), spread over the disk, amounting to about 25% by mass of the total molecular gas mass.

Thus a *first* a statistical analysis gets the mean of galactic longitude values for the different entries for a tracer t over each arm, and *second* a statistical analysis gets the mean offset in galactic longitude of the different means of tracer t for all arms – this 'linear sequence' aims to bring the errors down.

**Figure 4a** shows the mean positions (from a catalogue of arm tangents) for the mean CO tracer and the hot dust tracer, as a function of galactic longitude (Vallée 2016b). Adding 360$^o$ to the negative galactic longitudes (at left) would give a positive longitude value. One can readily see that the hot dust tracer of an arm is closer to the Galactic center than the mean CO tracer. Note the tracer *reversal* across the galactic center: at right ( l>0$^o$) the hot dust is at a smaller galactic longitude than the CO, while at left ( l<0$^o$) the hot dust is at a greater galactic longitude than the CO.

**Table 2** shows the statistics on the linear separations of each tracer, with 18 entries averaged over 6 half-arms. It confirms earlier efforts (Vallée 2014a, table 4; Vallée 2014c, table 2; Vallée 2016b, table 2). In Table 2, the first four arms are in Galactic quadrant IV, and the hot dust is separated from CO by a positive value towards the GC; the last row shows its mean separation from CO to be between 200 and 400 pc towards the GC. In Table 2, the last 2 arms are in Galactic quadrant I, and the hot dust is separated from CO by a positive value towards the GC; the last row shows its mean separation from CO to be between 300 and 400 pc towards the GC. Here we took account of the reversal of hot dust and CO between galactic quadrants IV and I.

In Table 2, the method to obtain the linear separation S is obtained simply by using the angular separation in galactic longitude between the CO arm tangent and the other tracer arm tangent in Column 1, then by multiplying this angular separation by the distance between the sun and the arm tangent. The mean separation S and its standard deviation of the mean (s.d.m.) are given in col. 8 and 9.

The data for each arm tracer are taken from the publication as published in the refereed literature (Vallée 2016b), and include:

-radio observations of HI atom at a wavelength of 21cm, radio observations of $NH_3$, $^{13}CO$, $^{12}CO$ molecules, radio observations of recombination lines at 1.4 GHz and of relativistic synchrotron emission at 408 MHz, radio observations of masers in starforming regions, and radio centrimetric observations of thermal free electron in the interstellar medium through pulsar dispersion measure at centimetric wavelengths,
-Far Infrared observations of cold dust,
-Mid-Infrared observations of [CII] and [NII] lines,
-Near Infrared observations of stars and of hot dust,
-Optical and radio observations of HII ionized regions,
-Gamma-ray observations at 1.8 MeV of the $^{26}Al$ (aluminum) atom.

**Figure 4b** shows the onion-like disposition of the different arm tracers inside a spiral arm, with the Galactic Center being the onion's center. The innermost onion sheet (red) represents the hot dust lane, due to the shock when the gas enters the inner edge of a spiral arm. The outermost onion sheet (blue) represents the cold, extended CO gas in the arm middle. The arm tangent are about 6 kpc from the Sun – see Table 1 in Vallée (2016b), not nearby. Hence a $2.5^o$ separation between CO and dust converts to 260 pc – see table 1 in Vallée (2016b). Inversely, a 200 pc separation at 6 kpc converts to 0.033 radian or $1.9^o$.

The peak of the old HII regions is at about the same place as the peak of the large angular scale $^{12}CO$ 1-0 gas emission, near the arm center. Visible old stars fill all the arms. Following Fig. 2 in Vallée 2014c, one divides the arm tracers into 4 zones:

-A blue zone is drawn for arm tracers at or very near the arm center, or mid-arm, around 0 pc, which include the cold $^{12}CO$ J=1-0 molecular gas as observed with a big half-power beamwidth of 8'. It encompasses tracers who fall nearby, and have a linear separation (from $^{12}CO$) smaller than their error (1 s.d.m.). Examples are the peak of the HII region complexes, the peak of old stars seen in NIR, and the peak of thermal electron, etc. This zone contains older evolved stars and older visible HII regions.

-A green zone is drawn for arm tracers further on, at around 100 pc away from the mid-arm, with arm tracers having a linear separation of that order, larger than their error (1 s.d.m.). Examples are the peak synchrotron radiation and radio recombination lines near H166α at 1.4 GHz. This zone might contain radio emission from young supernovae and young stars.

-An orange zone is drawn for arm tracers farther out, around 200 pc from the mid-arm, with arm tracers having a linear separation of that order, larger than their error (1 s.d.m.). Examples are the colder FIR dust and the masers. This zone might contain very young star-forming regions.

-A red zone is drawn for arm tracers way out, around 300 pc from the mid-arm, close to the arm's edge facing the direction to the Galactic Center (inner arm edge). Examples are the hot NIR and MIR dust emission. This zone might contain extended dust shocked by the galactic density wave.

The only theory that predicts such an offset for different tracers is the galactic density wave theory (Lin & Chu 1964; see also Roberts 1975). The arm cross-cuts are made between the Sun and the GC, hence at typical radial galactic distance from the GC of about 5 kpc. Inside corotation, with the orbiting gas going faster than the spiral pattern, the density wave theory predicts some offsets as follow: the hot dust lane (inner edge), the gas maximum and HII regions (arm center). Not all offsets shown in Fig. 4 have been predicted by the density wave theory.

4.2 Filaments

What is the exact location of the giant (100 pc) molecular filaments [GMF] - are they mostly in the interarm, or in the spiral arm ? If in the arm, are they in the dust lane, or do they fit the middle of the arm (spine)? Several narrow, long (50 pc to 230 pc) giant molecular filaments are known. Some authors place the long molecular filaments in the middle (center) of an arm. Thus in GQ IV, the 400-pc "Nessie" GMF appears to be a "bone" in the Scutum arm at $338^o$ and -38 km/s (one filament in Fig. 4 in Goodman et al, 2014), although along that longitude the radial velocity changes very rapidly with distance from the Sun (20 km/s over 1 kpc), making the true distance unclear if some of the filament velocity is not entirely due to galactic rotation. Thus for a 10 km/s random velocity, there is a 0.5 kpc distance move, making Nessie an interarm object. Also, Zucker et al (2015) found some filaments near the mid-arm structure. Here one must know exactly where is the inner arm edge and the mid-arm, as the two are about 300 pc apart (Fig. 4).

Yet some authors place the filaments at the inner edge of an arm. Wang et al (2016) located many cold filaments and found 1/4 of them to be close to an arm center. Ragan et al (2014) found the kinematical distance in a plot of radial velocity versus galactic longitude for many filaments to be in the interarm regions (seven filaments in Fig. 4 in Ragan et al 2014). Abreu-Vicente et al (2016) position about half of the filaments in the interarms (8 out of 17 in their fig. 1). For their GMF, their table 2 and Section 3.1.1 gave different kinematic and extinction distances (about 3.4 kpc versus about 4.9 kpc), or a difference about 3 times that of a typical arm width (≈0.6 kpc), making it difficult to say if a Giant Molecular Filament is located inside or outside an arm. Wang et al (2016) defined the location of a spiral arm in the radial velocity versus galactic longitude space (their figures 4a and 4b) and claimed that 27% of their long (100pc) filaments delineated the center of spiral arms (within an arm width of 10 km/s), and that a further 20% were interarm features (or spurs). Only more observations, with a more precise distance determination, could tell us exactly where a giant molecular filament fits inside or outside a spiral arm.

Theory-wise, numerical simulations, including self-gravity and feedback, showed how easy it is to reproduce giant molecular filaments (40pc to 250 pc in length) in the interarm regions through galactic shear, and these GMF appear to be situated either in the interarm or else in the process of joining an arm – see Duarte-Cabral & Dobbs (2016 – sect. 4.2.1).

4.3 Discussion

There is an *issue* about fitting locally and projecting globally – or how far can you extrapolate? All approaches use limited observations plus assumptions to derive a model. Many model fits to the spiral arms have only employed nearby tracers, such as optically visible open star clusters in the dusty galactic disk, hence within 3 kpc of the Sun's location. Without observing an arm tracer at large distances from the Sun, the fitted model parameters have large error bars, and their projection over the whole galactic disk is challenging (fig. 6 in Kharchenko et al 2012; fig.4 in Siebert et al 2012).

There is an *issue* about the existence or not of two superior arms. Is Sagittarius major or minor? Are there some arms more intense than others? The radio observations of Urquhart et al (2014 – their Sect. 4.1.3) found the Sagittarius arm to be prominent, strongly detected and clearly defined, traced by massive star formation; they concluded that the Sagittarius arm is a major feature of the Galaxy, rather than a minor arm. The near-infrared range is biased by massive extinction of light by dust, so these data must be calibrated as a function of longitude (Steiman-Cameron, 2010; Steiman-Cameron et al 2010 - their section 4.1). Some observers prefer the near-infrared range of wavelengths; they concluded that the Sagittarius arm is not a true arm, nor is the Norma arm a true arm (Benjamin 2008; Benjamin 2009). Once one does a proper dust correction, then other arms show up. Taking all wavelengths into account, the crosscut of alternating spiral arms shows all arm widths to be nearly equal (Vallée 2014a - his fig.3). A sizable minority of nearby spiral galaxies possess a 4-arm symmetry along with a bar, such as NGC 5970, NGC 0180, NGC 3346, NGC 6744.

There is an *issue* about the possible merging of the Perseus and Sagittarius arm. Some numerical models use only 2 spiral arms, along with bifurcations and mergings. Siebert et al (2012) derived a Sagittarius arm merging with the Perseus arm, at a location some 3 kpc away from the Sun near galactic l=90$^o$. The reality is that most spiral galaxies show a flat pitch angle with galactic radius (Honig & Reid 2015), hence if all arms start near the GC then merging of two major arms further away is not possible.

## 5. Where is the location of the start of each of the four spiral arms, near the Galactic Center?

In this section, the goal of this chapter is to provide the reader with the observed start of the Perseus arm near the Galactic Center, as well as that of the Scutum arm down there, and the recently observed start of the Sagittarius arm, and that of the Norma arm.

5.1 The start of the four spiral arms

The observed 'start of Perseus' arm, or Perseus arm origin, is already well documented, between galactic longitudes -23$^o$ (= 337$^o$) for the $^{12}$CO 1-0 peak intensity and -20$^o$ (=340$^o$) for the hot dust lane (Bloemen et al 1990; Bronfman 2008; Table 2 in Vallée 2008a).

Similarly, the observed beginning of the Scutum arm is well documented, between galactic longitudes +26$^o$ (dust lane and methanol masers) and +33$^o$ for the $^{12}$CO 1-0 peak intensity (Bloemen et al. 1990; Bronfman 1992; Table 1 in Vallée 2014c).

Concerning the other two spiral arms, an early attempt by Green et al (2011) using methanol masers (but not $^{12}$CO) suggested a link of the arms to the ends of the long thin bar (as known then). Also, the ends of the long thin bar nowadays cross some spiral arms, without apparent arm disruption.

One can accurately model spiral arms (see basic equations in Vallée 1995), using as input the best observations of arm tangents between the Sun and the Galactic Centre (GC), in a galactic radial range from 3 to 8 kpc, and also using the best observed spiral pitch angle value. The resulting output should be the best prediction for the start of these 4 arms near the GC (arm origins between 2 and 3 kpc). One finds that the tangent to the 'start of the Sagittarius' spiral arm (arm middle) is at l= -17$^o$ ±0.5$^o$, (=343$^o$) while the tangent to the 'start of the Norma' spiral arm (arm middle) is at l= +20$^o$ ±0.5$^o$. The origins of these four arms is found to be close to 2.2 kpc from the Galactic Center (Vallée 2016a).

**Figure 5** is a representation of the many additional features occurring near the Galactic Center (Vallée 2016a). The central galactic bulge (diamond, oriented at 25$^o$ to the Sun-Galactic Center line), and the blurred molecular ring near 3.7 kpc are also sketched. The short outflows (orange) represent the small +3kpc 'arm' (a misname short feature) and the small – 3kpc 'arm' (a misname short feature).

Recent observations of the Milky Way's central area shows three possible bars there, somewhat overlapping in space (the bulge bar, the nuclear bar, and the thin long bar).

Much of this describes recent observations and models that are not fully agreed upon and that are sometimes in conflict with each other.

In this figure, as seen from the Sun, one first finds the tangent to each arm, near their origin where the arm turns and becomes tangent to the line of sight from the Sun, and then one looks for observations of chemical tracers for this arm (and their offsets).   Observations of these chemical tracers showed that the tracer ordering ends exactly where the arm tangents are extrapolated from $^{12}$CO, giving an observed 'confluence' (between the tangent point and the ordered tracers: hot dust, masers, stars, etc).

In this figure, the observations showed, very near the arm tangents to 2 spiral arms, filaments going at high speeds (a misnamed '3-kpc arms' feature) which are potentially related to the spiral arms of Norma (at l=+20$^o$)and of Sagittarius (at l=-17$^o$). Misnamed '3-kpc arm' features far from the Galactic Center (at l>+13$^o$ or at l< 347$^o$) may be shocks associated with long spiral arms (Vallée 2017c).

5.2 Individual versus global pitch angles

**Table 3** shows for the Milky Way that each arm has a similar pitch angle, if one excludes the innermost and outermost regions (longitude 337$^o$ arm, and Cygnus +1 arm)..

The data in Table 3 for each arm is taken from the literature, using either the parallax method, the kinematic method, the luminosity-distance method, or the tangential method using the arms on both side of the sun-galactic center line. Most of the data used come from the parallax method employing masers.

Comparing each arm to the next arm, the last row shows that the mean individual pitch angle is often the same, from arm to arm, within the relative observational standard errors (about 3$^o$).

Elsewhere, in many nearby disk galaxies, the pitch angle appears roughly flat over 10 kpc, save for localized deviations with an amplitude near 20$^o$ (Honig & Reid 2015 - their fig. 9).

5.3 Central bars

There is a thick bulge bar (about 2.1-kpc radius and at about 20$^o$ to the line-of-sight – see Table 3 in [24]). The existence and nature of central bar(s) is an area of active work and debate. There are as well other 'putative' bars (one is a 4.2-kpc-radius bar at about 40$^o$ to the line-of-sight – see Table 3 in [24]; and one is a 0.5-kpc-radius bar almost perpendicular to the line-of-sight – see Table 2 in [24]).

Is the long bar a composite of segments, as it appears to cross four spiral arms (Fig.5 here) without any observable effect? Some groups view the long bar as a composite: a bar and an S-shaped trail (fig. 17 in Wegg et al 2015), or else a physical link between the outer edges of the short bulge bar and those of the long bar, appearing as 'leading ends' (Fig. 1 in Martinez-Valpuesta & Gerhard 2011), while Monari et al (2017) have argued for the long bar being a loosely wound spiral, not a straight structure.

5.4  Discussion

As mentioned in the Introduction, the molecular ring is a rather blurred composite of various segments, and is regarded by some as not a physical barrier to the arms – see the discussion in Vallée (2008a, section 3.2).  With a higher angular resolution, the molecules were found in specific areas, not in a whole ring. Thus once one removes the origins of each of four arms in that radial range, then very little gas appears to be left there in velocity space (Dobbs & Burkert 2012).

There is still an *issue* about the location of the so-called misnamed '3-kpc arm' features. They perhaps should be relabeled '3-kpc-streams'. Some have been assigned close to the well-known 'origin of the Perseus' arm, while others have been assigned to the well-known tangents to the inner Scutum arm. Others found the so-called '3-kpc arm' feature at various places, over a wide area near the Galactic Center between $12°<l<24°$ and $-12°<l<-21°$ (Vallée 2016a – his Table 5 and Section 7). These '3-kpc-streams' exist observationally, yet their ages and stability are unknown. They could be due to temporary turbulence, as sometimes predicted in some models (Vallée 2017c).

**6. Some explanations for the arm's orbital dynamics, formation, maintenance, and waves**

In this section, we provide the reader with basic observables for angular rotation of various galactic features, and an attempt at basic explanations of the observed spiral arms through the numerical or analytical theories for spiral arms.

6.1 Angular rotation and orbits
Starting from galactic radius R and orbital velocity V, one can obtain angular rotation ω and orbital period P, as follows:

$$R_{(kpc)} \cdot \omega_{(km/s/kpc)} \approx V_{(km/s)} \tag{6}$$

$$2 \cdot \pi \cdot R_{(pc)} \approx V_{(km/s)} \cdot P_{(Myr)} \tag{7}$$

where: 1 Myr = $3.16 \times 10^{13}$ s         and:    1 pc = $3.09 \times 10^{13}$ km
Note the approximative sign ($\approx$), as this is not a strict equality.

**Table 4** shows the results. The first column indicates the feature whose distance (column 2) angular rotation (column 4), circular velocity (col. 6) and orbital period (col.8) are given, along with their respective references. (col. 3, 5, 7, 9).
The 'spiral pattern' rows refer to the 'pattern speed of the spiral density wave' at the Sun's location (set at 8.0 kpc), while various published models (column 5) found it an angular velocity value (column 4), from which equation 6 here yielded its orbital speed at this radius (velocity in column 6), while equation 7 yielded an orbital period (column 8). The word 'circular' is mathematical, as the model-dependent pattern orbit may not be circular. This places the Sun just inside the corotation radius.
The 'bulge bar' rows refer to the orbital speed at the end of the short thick bar (set at a radius of 2.1 kpc [24]) straddling the stellar bulge, while various published models (column 5) found it an angular velocity value (column 4), from which equation 6 here yielded its orbital speed at this radius (velocity in column 6), while equation 7 yielded an orbital period (column 8). The word 'circular' is mathematical, as the model-dependent bar orbit may not be circular.

For the stars and gas near the Sun in the Local Standard of Rest (LSR), one can use the galactic distance of the Sun to the Galactic Center $R_{sun}$ (8.0 kpc) and orbital velocity $V_{lsr}$ to compute here ω and P for the stars and gas.

For the hot gaseous halo around the Milky Way (within 50 kpc) at a temperature near $2 \times 10^6$ K, one can use at a radius of 8.0 kpc the observed rotation velocity $183 \pm 41$ km/s around the Galactic Center (Hodges-Kluck et al 2016).

For the short boxy bulge bar at the Galactic Center, one can use the radius R = 2.1 kpc for the boxy bulge bar. One takes for ω the value predicted by the bar model.

Are there links between these ω values?

In the density wave theory, the LSR's angular value near the Sun ought to be slightly larger than the value for the spiral pattern, in order to provide a shock upon entering an arm (before co-rotation). Orbital streamlines around the Galactic Center predict that the orbiting gas will slow down as they approach a spiral arm, and speed up afterwards when leaving the arm.

It is noted that the ω value for the hot halo gas is close to but slightly smaller than the ω value for the spiral pattern, although there is no obvious direct explanation. Hodges-Kluck et al (2016) suggested that outflows from the disk into the halo may follow galactic rotation but eventually lag behind the gas in the disk. Expectations for the bulge bar vary. Englmaier & Gerhard (1999) and others argued that the angular rotation of the spiral pattern (≈ 25 km/s/kpc) ought to be similar to the angular rotation of the short boxy bulge bar (≈ 35 km/s/kpc), if the two features are *attached*. Bissantz et al (2003) used smoothed-particles hydrodynamics (SPH) to follow gas flows orbiting around the Galactic Center. Their models indicated the need for different angular speeds, the one for the bar differing from one for the spiral pattern, and sufficiently different to avoid a dissolution of the arms.

6.2 Formation and maintenance and waves

How do arms form in the Milky Way? There are a plethora of theories and simulations to generate spiral arms. Their predictions should match a few observed arm parameters: linear separation of different arm tracers from $^{12}$CO (over 350 pc), all arm tracers to one side of $^{12}$CO, the number of arms (4), average pitch angle (13° inward), arm shape (logarithmic) and azimuthal arm spacing (equal), to first order.

No small improvements in *distant*-based (> 30 kpc), tidal theories for the arm formation could generate this arm number (4), and the observed tracer offsets (red, orange, green, blue in Fig. 4b). No small improvements in *local*-based (< 3 kpc), individual random perturbing instabilities could show this equal arm ordering and tracer offsets. Perhaps additions in *galactic*-based (≈ 10 kpc) wave theories and simulations might be made to explain the ordered offsets for different chemical tracers in four equidistant spiral arms.

6.2.1 Versions of a *globally*-based theory and simulations

Theories having shocks can create offsets among arm tracers. The quasi-stationary density-wave theory can produce several arms, an orbital trajectory of the gas and stars going through spiral arms, and a linear separation between arm tracers as occasioned by a shocked gas and dust entering the arm from the inner edge, then some of it coalescing to form stars there (dust lane), progressing through the arm and exiting on the arm's outer edge (Roberts 1975 - fig.3). In Figure 5, one can assume that the corotation radius is above 9 kpc. Beyond corotation, with the orbiting gas slower than the spiral pattern, the wave theory predicts a reversal of the tracer offsets with respect to the arm center.

In the non-linear density wave, the *shock* position and the gas density maximum (at mid-arm) are *separated* by about 12 Myrs (Roberts 1975 - Fig.2), or 306 pc for a gas velocity difference of 25 km/s (Dobbs & Baba 2014 - fig. 17b); this separation between the shocked dust lane and the mid-arm is similar to the observed arm's halfwidth of 360 pc (Figure 2 in Vallée 2014c). The theoretical model with 4 arms of Gittins and Clarke (2004 - fig.16) has the gas density maximum (potential minimum, broad CO at mid-arm) separated from the shock lane (hot dust).

The density wave proposed that the peak of each arm tracer is *between* the inner edge of an arm and the mid-arm, excepting the old stars being distributed all over the arm. All arm tracers are to be found between these two boundaries, *none outside*; this is exactly what is observed in arm tangent catalogues (Fig.3 in Vallée 2014c).

The density wave predicts that the inner arm edge with the dusty shock lane shall be at a *higher* galactic longitude than the mid-arm in Quadrant IV, but at a *lower* galactic longitude than the mid-arm in Quadrant I – as observed in arm-tangent catalogues (Table 1 in Vallée 2014c).

Renaud et al (2013) used hydrodynamical simulations at a high resolution of 0.05pc, with stellar feedback, a bar, and many arms, finding a global molecular structure in the Milky Way disk. Their arms displayed a gradient of gas density and velocity across a spiral arm (their fig.13), and they found 3 or 4 spiral arms close to the bar (their Fig.17). They also find many 'beads on a string' with regular spacing along the arm.

6.2.2 Versions of *locally*-based theories

The transient-recurrent 'dynamic spiral' reconnection theory (Dobbs & Baba 2014, Sect. 2.2) has arms breaking and reconnecting, and a high pitch angle between $20^o$ and $40^o$ (Dobbs & Baba 2014, fig.10), unlike the mid $13^o$ pitch as found generally in the Milky Way. Baba et al (2016) proposed a dynamic spiral theory, in which the gas does not flow through the spiral arm (contrary to the density wave theory), but falls into the arm from both inner and outer sides (their Fig.5), and there is no offset between the gas density peak and the stellar spiral arm. One would need a shock on the inner side of the arm (but not on the outer arm side).

The stochastic, self-propagating star formation theory could produce a very irregular, flocculent arm pattern (Dobbs & Baba 2014, sect. 2.5), not the quasi-regular pattern seen in the Milky Way.

The swing-amplification theory could produce numerous flocculent arms irregularly located (Dobbs & Baba 2014, sect. 5.2), not quasi-regularly spaced as in the Milky Way.

6.2.3 Versions of bar-driven or *tidally*-based theories

The bar-driven theory could induce two strong arms (Dobbs & Baba 2014, sect.2.3), not four as found in the Milky Way. Perhaps each of two straddling galactic bars could perhaps produce its own set of twin arms.

The tidal theory involving a recent passage of the Sagittarius galaxy could induce two strong arms (Dobbs & Baba 2014, sect. 5.3), not four as found in the Milky Way. Perhaps the two Magellanic Clouds could produce two sets of twin arms.

6.3 Discussion

There is an *issue* about the number of different spiral density waves that could be co-existing. Most models assume 2 arms (m=2 waves) or 4 arms (m=4 waves). Griv et al (2015) assumed a collection of spiral waves, including m=1, m=2, m=3, and m=4 arms [95], leading to many arms. To deal with this collection, they employ a superposition of Fourier harmonics and azimuthal wave number m (= number of spiral arms for a given harmonic). To find which m is best, they employed a weight sum S of the velocity difference (observed minus modelled); S is also called later an 'estimator'. Their best fit to the data employs an 'estimator' S that yield similar values (between 24.4 and 24.9 for open clusters in their Table 1; between 4 and 5 for masers in their Table 2), irrespective of having no spiral (m=0) being fitted or of any m value being fitted (m=1 to 4). They suggested "a number of discrete spiral modes of collective oscillations" (their abstract). Figure 5 in Griv et al (2015) showed their preferred m=1 spiral mode. Their observed data are restricted to a short distance from the Sun (their equation 4 in Griv et al 2015). This small region encompasses a local 'Orion' spur in which the Sun is located, and they do not delete the data from this local spur, thus projecting its effects all around the whole Galaxy.

There is an *issue* about whether there are rings, and not spiral arms. Melnik et al (2016) assumed that there are two rings in the Milky Way, with the Sun as set at a galactic radius of 7.5 kpc, an outer ring R2 above the Sun at a mean galactic radius of 8 kpc, and the other ring R1 below the Sun at a mean galactic radius of 6 kpc (their Fig.1). The rings would be elliptically shaped (their Section 3). Their model tangent, as seen from the Sun, for the 'ascending' segment of R1 (Carina segment) is at a galactic longitude of +302$^o$ (their fig. 2b and fig. 4a). This is in between the observations of the arm tangent for the Carina arm near l= 283$^o$ and the tangent to the observed Crux-Centaurus arm near l= 310$^o$ (see Table 1 in Vallée 2014c). Their ring R1 is made up of an 'ascending' segment of pitch +6$^o$ (near the Carina arm, their Fig. 2b and Fig. 4a) and a 'descending' segment of pitch -6$^o$ (near the Sagittarius arm – their Fig. 2b and Fig.4a), giving a mean ring pitch of 0$^o$. But the observations show the Carina-Sagittarius arm to be a continuous spiral with the same pitch of -13$^o$ (Table 1, Figure 1, Section 1 in Vallée 2015).

Some numerical and analytical models use an original 2-arm model, and with some complex dynamics, they end up with a 4-arm gas response after a short while (after 300 Myears; beyond 3 kpc galactic radius). In a 2-arm structure close to the Galactic Center, each inner arm would develop a fork-like bifurcation producing 2 gaseous arms beyond. For these paired arms, one arm has a large pitch angle (near 15$^o$) and the other arm has a small pitch angle (near 7$^o$) as in Fig. 7 in Pérez-Villegas et al (2015). Thus when travelling on an orbit at a galactic radius of 8 kpc around the Galactic Center, one would meet successively 4 gaseous arms, alternating in pitch angle: one with a large pitch, then a small pitch, a large pitch, and a small pitch. This predicted alternation in pitch angle (along a circular orbit) is not seen observationally for the Milky Way – see Table 3. Also, their azimuthal angular separation, along the circular orbit, between the four arms would also alternate, from small (≈40$^o$ between the 2 arms of a single pair), to large (≈140$^o$), to small (≈40$^o$ between the 2 arms of the other pair), to large (≈140$^o$). Finally, the predicted large and small angular spacings (along a circular orbit) between adjacent arms is not seen in the Milky Way, as the longitude tangents change monotonically in longitude – see Fig. 4a here.

A Bayesian computer routine was proposed to model spiral arms, fitted to water and methanol masers. Their Carina arm has a computed arm tangent at longitude 302$^o$ (Fig.1 in Ragan et al 2016), contradicting published observations in a range from longitude 281$^o$ (CO) to 285$^o$ (dust) – see Vallée (2014a – table 3); this discrepancy is significant (17$^o$ at 5 kpc from the Sun is 1 450 pc). Also, their arm tangent to Norma is seen at galactic longitude l= 023$^o$, contradicting published observations of the arm tangents to the Norma between l=16$^o$ (masers) and 20$^o$ (CO) - see Vallée (2016b – table 3); this discrepancy is significant (4$^o$ at 6.5 kpc from the Sun is 454 pc). Finally, maser data can also be found in numerous small 'spurs' or 'armlets', and if not taken out they could be falsely taken as a part of long spiral arms (as defined with a dozen of other tracers).

### 7. Other potential players, in the galactic disk

In this section, the aim of this chapter is to provide a summary of some of the remaining galactic players that could affect the galactic disk, yet have not been observed so far to be at a high significance level. These remaining players may be independent, or may be linked; other players may yet to be identified in the future.

#### 7.1 Magnetic field map and relations to the arms

Magnetic fields in the Milky Way's disk were recently reviewed by Beck (2016), Haverkorn (2015), Heiles & Haverkorn (2012), Van Eck et al (2011), Vallée (2012), Vallée (2011). Basically, there are large radial regions where the magnetic field direction is going clockwise (between 2 and 5 kpc from the Galactic Center; and between 7 and 12 kpc from the Galactic Center), or counterclockwise (between 5 and 7 kpc from the Galactic Center) – see Vallée (2011, fig.30).

The electro-magnetic waves passing through the interstellar medium of our Galaxy are affected by the magneto-ionic medium in the galactic disk, as the wave's polarized electric vector is rotated by the Faraday Rotation. Magnetic fields are deduced after the fitting of a Faraday Rotation Measure (RM) to the observed wavelength-dependent angles of the electric vectors of a source's linear polarization emission.

Finding and eliminating an ambiguity of 180$^o$ in the position angle of the electric vector of polarization is important. The Faraday Rotation Measure Synthesis technique is fraught with this problem. Thus some results Braun et al (2010) and Heald et al (2009) on nearby spiral galaxies have claimed that a galaxy can have 3 galactic disks, each one on top of the other – a middle one full of stars and gas, and 2 pseudo-disks without stars and located on each side of the middle one; later it was shown that this is possibly due to the non-removal of the 180$^o$-ambiguity mentioned above (Vallée, 2013b, fig.4; Mao, 2014, Fig.5).

#### 7.1.1 Magnetic field reversal.

**Figure 6** shows a representation of the general magnetic field lines and magnetic field directions within the disk of the Milky Way. Adapted from Vallée (2008b). as obtained with observed pulsar RM. The global disk magnetic field is essentially clockwise (blue arrows), except for a small zone where the magnetic field has reversed (red arrows), and has a small inward pitch angle.

There is a small magnetic inclination inward (a small pitch angle, inward from the circle).

Van Eck et al (2011) used the Faraday rotation measures of extragalactic point sources (quasars or elliptical galaxies) to infer in our interstellar medium an AxiSymmetric spiral magnetic field, with a single reversal zone (their Fig.11). Their model is very similar to that in Vallée (2008b - fig.5), differing only very near the Galactic bar where RM data are few.

The recent review of the Milky Way's disk magnetic field, by Beck (2016, his Fig. 26), is also very close to that in Vallée (2008b - fig.5).

Smoothed-particle magneto-hydrodynamics simulations with an imposed spiral potential can reproduce a magnetic reversal at a radius of around 4 to 6 kpc in the Milky Way (Dobbs et al 2016).

Is the magnetic field attached to each arm? Some early models of the magnetic field direction alleged that the magnetic field was directed counterclockwise inside each spiral arm, and clockwise outside each spiral arm (in the interarm) – see Han et al (2006, fig.12) and Han (2013). With 4 arms, one should see eight magnetic field reversals - the reality is that one does not observe so many magnetic reversals in other spiral galaxies, and neither in the Milky Way.

### 7.1.2 Magnetic field pitch.

Kronberg & Newton-McGee (2011) found that the pitch angle of the magnetic field lines (outside spiral arms) is close to $-5.5^o \pm 1^o$, thus a bit different than the pitch angle of gas, molecules and dust inside spiral arms at $-13.1^o$, as found observationally in many arm tracers.

In nearby spiral galaxies, Van Eck et al (2015) found a small difference between the pitch angle of the magnetic field and the pitch angle of the stellar arm, with the magnetic pitch angle being more open (larger) than that of the stellar arm, by about $5^o$ to $10^o$ (their sect. 5.2).

The early *issue* about a magnetic control of a spiral arm has gone away. Some early models employed a strong magnetic field, enough to control the arm. In this case, the pitch angle of the magnetic field would be the same as the pitch angle of the stellar arm.

The basic density wave theory does predict a small difference in pitch angle between the interarm magnetic field direction and arm direction (Roberts 1975 - his Fig.3). Assuming a 4-arm model in the Galactic disk, the orbital path of gas and magnetic field around the Galactic Center becomes a continuous series of 4 segmented spirals (Vallée 2011 - fig.37). When the orbit of a gas and dust patch enters a spiral arm, the gas is shocked and the dust is heated, and the magnetic field line is reoriented.

### 7.2 Dark Matter map and relations to the arms

Is there a dark matter disk? Since most mergers of nearby dwarf galaxies with the Milky Way end up in the Milky Way disk, some dark matter and newly accreted stars from these dwarfs should be found also in the Milky Way's disk. Bregman (2012) argued for a model where unobservable baryons are in the halo, because either they were heated and expelled from the disk by supernovae or galactic superwinds, or they were heated early in the halo and thus resisted the infall on the disk and were later pushed in the intracluster medium.

McKee et al (2015) reanalyzed the observed data in the solar neighbourhood (stars, gas and baryons) to find the local density of all matter at about $0.10 \pm 0.01$ $M_{sun}/pc^3$; this value comprises a dark matter component at about $0.013 \pm 0.003$ $M_{sun}/pc^3$ (about 13% of all matter), but there is no indication of a vertical structure (in the plane or homogeneous) or of an horizontal structure (in an arm or an interarm). Fig. 1 in Xia et al (2016) showed that the median local dark matter density for 6 measurements between 2013 and 2016 to be $0.014 \pm 0.004$ $M_{sun}/pc^3$ (about 14% of all matter). Independently, using the Sloan Digital Sky Survey with about 10 000 stars, Read (2014) finds a rather weak 'dark matter disc', amounting to about 10% of the total mass in the local volume ($8<R_{sun}<9$ kpc; $0.5$ kpc $< z < 0.5$ kpc). McGaugh (2016) found a solar neighborhood local dark matter density of $0.009$ $M_{sun}/pc^3$ (her equ. 9), or about 9% of the total matter density. Using the Gaia-ESO Spectroscopic Survey with 4 600 stars, the observed absence of accreted stars in the disk suggests an absence near the Sun of dark matter in the disk (the 'dark disk' of Ruchti et al 2015).

It is noteworthy that the mean dark matter density from the above values (from 9% to 14%) is not higher than a 3-sigma result.

## 8. Conclusion, Summary, and Outlook

In summary, a review of the basic observed spiral arms was made (Figures 1, 2, 3 here). In the latest maps of the Milky Way galaxy, each spiral arm has separated into slightly different parallel lanes (onion like), each with different chemical components (arm tracers). The presence of offsets among the arm tracers in the Milky Way (Figure 4a, and Figure 4b) confirms each arm as a true arm, and thus disfavors any 2-arm model. A summary of what's close to the Galactic Center, encompassing the start of the spiral arms, is given (Figure 5). A recent model magnetic field picture is shown (Fig. 6).

In Section 2, the reader has been provided with a global view of the Galactic disk domain, summarizing numerous papers with respect to the locations of spiral arms, their shape, pitch, separation, how to subtract foreground and background emissions in order to get the data for a given spiral arm. The Sun is located in a 'spur' in an interarm, between the Sagittarius and the Perseus arm. Another 'spur' has been observed in an interarm, around longitude $35^o$ and radial velocity near +75 km/s, somewhere between the Scutum arm and the Sagittarius arm.

In Section 3, the reader has been provided with a radial velocity view of the gas in each spiral arm, located along the galactic disk. Older, published cartographic maps of the spiral arms showing weird 'fingers of God' pointing to the Sun were shown through models to be velocity jumps (not linked to orbital velocity) masquerading as pseudo physical jumps (fingers).

In Section 4, arm tangents and arm widths, with the mean galactic longitude at which each arm tracer, chemical or otherwise, are summarized. Always, for each spiral arm, the arm tangent for the hot dust (inner arm edge) is closer to the Galactic Meridian (at galactic longitude $0^o$) than the arm tangent for the $^{12}CO$ tracer (middle arm position). This shows the Galactic Meridian to be some kind of 'axis of reversal' with respect to arm tracers.

In Section 5, the reader is provided with a view of the galactic interior, the inner area within about 3 kpc of the Galactic Center (Vallée 2017d). A discussion there touched on the existence of an inner molecular ring, near a galactic radius of 3.5 kpc, and on the existence of a long thin physical bar crossing some spiral arms, as well as the '3-kpc-arm' features that could be physical streams (not arms).

In Section 6, the reader is appraised of angular velocities and potential matches between observational data and model predicted features and numerical simulations, including shock locations, tracer offsets, HII peak. Some numerical simulations are compared to the overall observational data.

In Section 7, some other potential players in the galactic disk were briefly reviewed. The observed weak strength of the global disk magnetic field, and the observed weakness of the dark matter in the galactic disk, make them both minor players in maintaining the spiral arms. The pitch angle of the magnetic field in the arm is observed to be slightly different than that of the stellar arms, as predicted by theories involving magnetic fields.

Table 1 lists recent observational efforts and results to find the exact distance from the Sun to the GC, and the circular velocity of the local stars around the GC.

Table 2, listing the observed linear separation of each specific arm tracer from the $^{12}CO$ tracer, in each spiral arms, indicates a notable substructure inside an arm. With the discovery of substructure in spiral arms, reversing across the Galactic Meridian, the Milky Way is pointing us to a specific formation model for its spiral arms, with a very specialized onion-like ordering.

Table 3 details the pitch angle of each spiral arm, as observed by different methods. Each method has some assumptions, and covers a different portion of each arm. Hence statistics can be used to obtain a more global picture.

Table 4, listing the angular rotation of some of the important Milky Way features, indicates some possible links. Unfortunately, the error bars are currently not small enough to eliminate many theories on these topics.

Still to be determined are the locations and relations between the 100-pc giant molecular filaments and the nearby spiral arms. Are they near the inner arm edge or in the interarm as predicted by some theories, or do they form the 'backbone' at the center of a spiral arm? Their exact distances from the Sun is not well known enough, but would have to be more precise than a half-arm width (340 pc).

**Acknowledgements.**
The figure production made use of the PGPLOT software at the NRC Canada's Herzberg A&A in Victoria, BC. I thank two anonymous referees for useful, careful, and historical suggestions.

# References.

**Table 1 – Recent measures of global parameters of the Milky Way**[a]

| $R_o$ (kpc) | $V_{lsr}$ [b] (km/s) | Data used | References |
|---|---|---|---|
| 8.3 | - | stellar orbits near GC | Gillessen et al (2017 – Sect. 3.6) |
| 8.2 | 233 | main mass model | McMillan (2017 – Table 2) |
| 8.0 | 227 | alternate mass model | McMillan (2017 – Table 6) |
| 7.6 | - | GKM stars | Branham (2017 – Table 3) |
| - | 231 | Cepheids | Bobylev (2017- Fig.2) |
| - | 219 | OB stars | Bobylev & Bajkova (2017 – Sect.8) |
| 8.34 | 240 | masers | Reid et al [125] |
| 8.1 | 230 | Mean value | |

Note:
(a): For earlier data, see Vallée (2017a).
(b): The kinematic circular rotation value near the Sun, excluding the Sun's peculiar velocity.

# Table 2 - Linear separation (S) of each arm tracer from $^{12}$CO, in each spiral arm[a]

| Chemical Tracer | S Carina arm | S Crux-Centaurus arm | S Norma arm | S Start of Perseus arm | S Scutum arm | S Sagittarius arm | Mean separation <S> | s.d.m.[b] |
|---|---|---|---|---|---|---|---|---|
| | pc | pc | pc | pc | pc | pc | pc | pc |
| Galactic Quadrant | IV | IV | IV | IV | I | I | | |
| $^{12}$CO at 8' | 0 | 0 | 0 | 0 | 0 | 0 | 0 | _[c] |
| Old stars (NIR) | - | -209 | - | 209 | 26 | -314 | -72 | 118 |
| [CII] at 80" | - | - | -49 | 28 | - | - | -10 | 38 |
| $^{26}$Al | - | 52 | -415 | - | 78 | 316 | 8 | 137 |
| Thermal electron | 148 | -52 | -49 | - | 78 | 105 | 46 | 41 |
| HI atom | 22 | 42 | -37 | 0 | 166 | -7 | 31 | 29 |
| NH$_3$ 1-1 cores | - | -42 | -73 | 223 | - | - | 36 | 94 |
| HII complex | 218 | 42 | -244 | 56 | 288 | -7 | 59 | 76 |
| $^{13}$CO | - | - | - | - | 140 | 21 | 80 | 59 |
| Synch. 408 MHz | - | 52 | -49 | 307 | 78 | 175 | 113 | 60 |
| 1.4GHz RRL | 262 | 178 | 110 | 14 | 183 | 91 | 140 | 35 |
| Warm $^{12}$CO cores | - | - | - | - | 253 | 105 | 179 | 73 |
| FIR [CII] & [NII] at 7° | 497 | -52 | - | 167 | 253 | 35 | 180 | 95 |
| Cold dust 870µm | 253 | 198 | 147 | 140 | 174 | 98 | 168 | 22 |
| Cold dust 240µm | 235 | 157 | 440 | - | 166 | 35 | 207 | 67 |
| Combined cold dust [d] | 244 | 178 | 294 | 140 | 170 | 66 | 182 | 33 |
| Masers | - | - | 244 | 140 | 384 | 35 | 201 | 74 |
| Hot dust 60µm | 322 | 157 | 73 | 446 | 602 | - | 320 | 96 |
| Hot dust 2.4µm | - | - | 449 | 307 | 340 | - | 365 | 43 |
| Combined hot dust [e] | 322 | 157 | 261 | 377 | 471 | - | 318 | 53 |

Notes:
(a): Data from Table 3 in Vallée (2016b).
(b): The s.d.m. in the last column is from the external scatter.
(c): There is a median internal scatter of 40 pc, from the CO data (Vallée 2016b – table 5).
(d): Statistical means made on both cold dust tracers at 240µm and 870µm.
(e): Statistical means made on both hot dust tracers at 2.4µm and 60µm.

**Table 3 – Observed individual pitch angle (p, in degrees, negative inward), for each spiral arm in the Milky Way galaxy[(a)]**

| Norma arm | Scutum-Crux-Centaurus arm | Carina-Sagitt. arm | Perseus arm | Cygnus arm | Cygnus +1 arm | | | |
|---|---|---|---|---|---|---|---|---|
| l=329° | l=309° | l=281° | - | - | - | | | |
| l= 20° | l= 33° | l= 51° | - | - | - | | | |
| p | p | p | p | p | p | Method[(b)] | Data used[(c)] | Reference |
| - | - | -19.0 | - | - | - | par | meth. masers | Sect.6.2 in Krishnan et al (2017) |
| -13.7 | - | - | - | - | - | tan | CO, masers | Equation 4 and below (this review) |
| - | - | - | - | -13.1 | - | kin | CO gas | Fig.2 in Du et al (2016) |
| -15 | -11 | -11 | -15 | -15 | - | kin | HI and CO | Tab. 1 in Nakanishi & Sofue (2016) |
| - | -13.3 | -14.0 | - | - | - | tan | CO, HII, dust | Tab.1 and 2 in Vallée (2015) |
| - | -19.2 | - | - | - | - | par | meth. masers | Sect.5.2 in Krishnan et al (2015) |
| - | - | - | -11.1 | - | - | par | $H_2O$ masers | Fig.4 in Sakai et al (2015) - VERA |
| - | -10.0 | -10.5 | -7.9 | -10.3 | - | lum | Cepheids | Table 1 in Dambis et al (2015) |
| - | - | - | - | - | -9.3 | kin | CO | Fig.3 in Sun et al (2015) |
| - | - | - | - | -14.9 | - | par | $H_2O$, meth. masers | Fig. 6 in Hachisuka et al (2015) |
| | | | | | | | | |
| -14 | -12 | -13 | -11 | -14 | - | | Median value | |
| -14 | -13 | -14 | -11 | -13 | - | | Mean value | |

Notes:
(a): For earlier data, see Vallée (2015).
(b): Kinematic method (kin), Parallax method (par), Luminosity-period Cepheid method (lum), or Twin-arm tangents method (tan)
(c): GMC = Giant Molecular Clouds; HII = HII regions; meth. = methanol

**Table 4– Recent angular rotation values of some Milky Way galactic features**

| Feature [a] | R gal. radius (kpc) | ref. [b] | $\omega$ ang. vel. (km/s/kpc) | ref. [b] | V circ. vel. (km/s) | ref. | P orbital period (Myr) | ref. |
|---|---|---|---|---|---|---|---|---|
| - - - - - - - - - - - - - - - - - - - - - - - - - - - - - - - - - - - - - - - - - - - - - | | | | | | | | |
| LSR [a] | 8.0 | set | 28 | equ.6 | 220 | set | 228 | equ.7 |
|  | 8.0 | set | 28.7 | N11 | 230 | equ.6 | 219 | equ.7 |
|  | 8.0 | set | 28.8 | equ.6 | 230 | set | 220 | equ.7 |
| Spiral pattern | 8.0 | set | 25 | D15; G11 | 200 | equ.6 | 250 | equ.7 |
|  | 8.0 | set | 23 | L16, J15 | 184 | equ.6 | 273 | equ.7 |
|  | 8.0 | set | 20 | K16 | 160 | equ.6 | 314 | equ.7 |
| Bulge bar | 2.1 | V16a | 30 | R08 | 63 | equ.6 | 209 | equ.7 |
|  | 2.1 | V16a | 40 | Q15 | 84 | equ.6 | 157 | equ.7 |
| - - - - - - - - - - - - - - - - - - - - - - - - - - - - - - - - - - - - - - - - - - - - - | | | | | | | | |

Notes:
  (a) LSR = Local Standard of Rest, surrounding the Sun's position in the Milky Way disk.
  (b) D15 = Dambis et al (2015); G11 = Gerhard (2011 – Sect.4); J15 = Junqueira et al (2015 – tab.4); K16 = Koda et al (2016 – Tab.2); L16 = Li et al (2016 – sect. 2.2.3); N11 = Nagayama et al (2011 – fig.6b); Q15 = Qin et al (2015 – Sect. 2); R08 = Rodriguez-Fernandez & Combes (2008 – Sect. 6.1); V16a = Vallée (2016a – tab.3)

**Figure captions:**

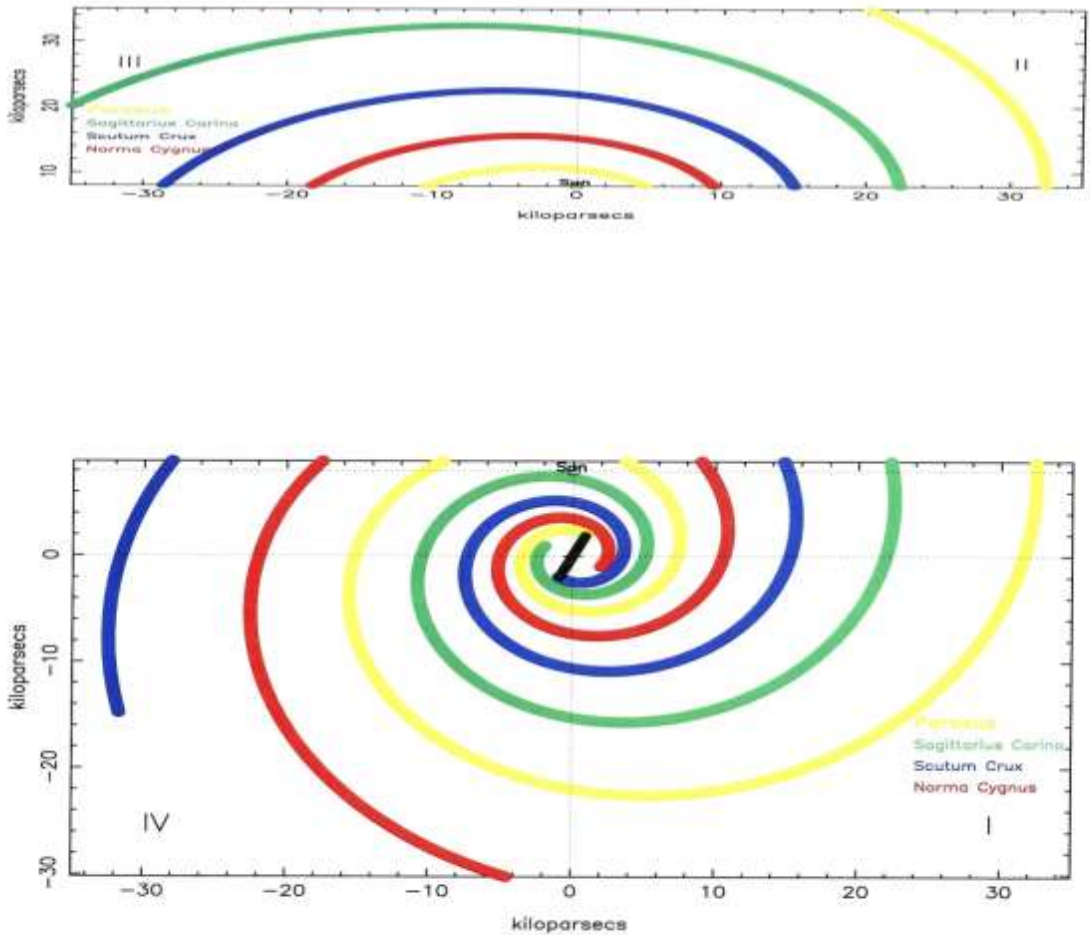

**Figure 1.** Sketch of the location of each spiral arm, seen at different galactic longitudes. Input: 4 arm, log shape, pitch angle of 13.1° inward, arm start at 2.2 kpc, distance Sun to Galactic Center of 8.0 kpc. The Sun is shown by a circle. Each spiral arm is colored differently. Adapted and updated from Fig.2 in Vallée (2016a).
  (a) Toward the Galactic Anticentre. The Sun is at the bottom center. Galactic quadrants III (left) and II (right) are shown.
  (b) Towards the Galactic Center. Galactic quadrants IV (left) and I (right) are shown. The Galactic Center is shown by a cross at (0,0). The part of an arm facing the Galactic Center is called the 'inner edge', while across the arm the part facing the Galactic Anticenter is called the 'outer edge'. The arm sketch for Scutum-Crux and the one for Sagittarius-Carina have two arm tangents as seen from the Sun (located in Quadrant IV and Quadrant I).

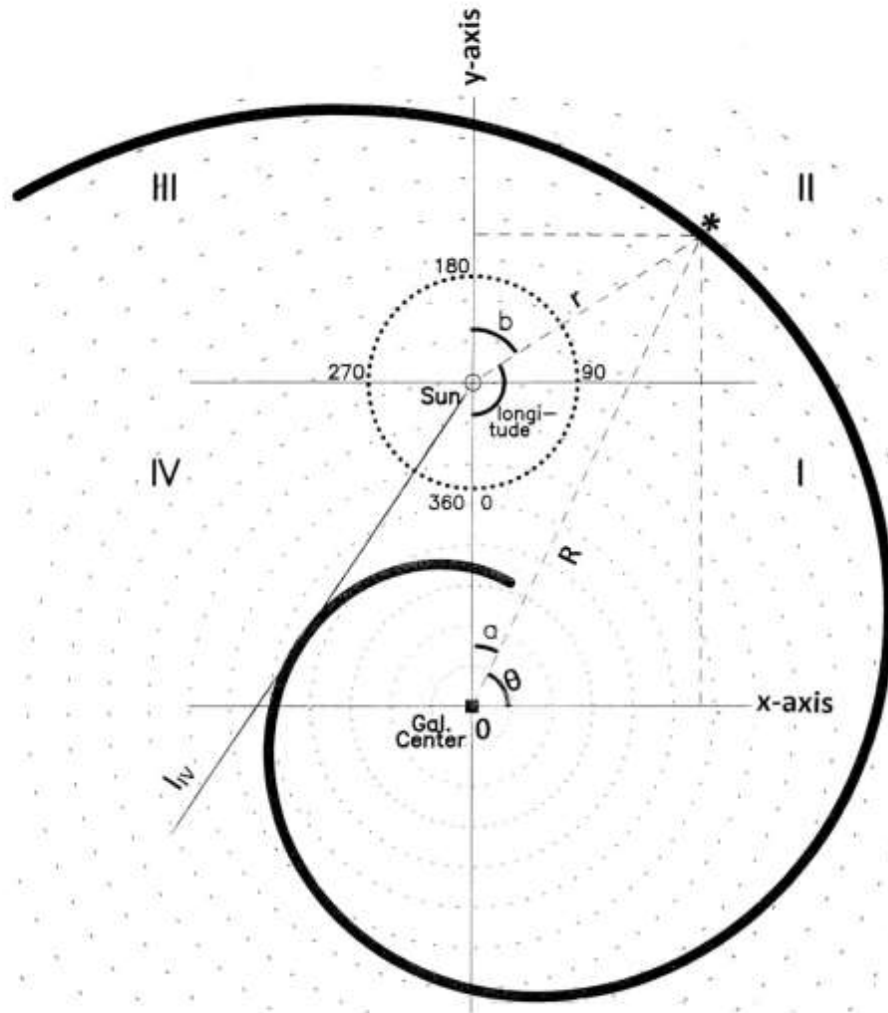

**Figure 2**. Sketch of the twin coordinate system employed. The Sun (small circle) is at 8.0 kpc from the Galactic Center (small square). A median circle (thick dots) around the Sun shows the Galactic longitudes. A series of large circles (dashes) around the Galactic Center shows the circular orbits of gas and stars. An object (black star) on the spiral arm is shown at angle θ and distance R from the Galactic Center, thus at longitude l and distance r from the Sun. This sketch approximates the Norma-Cygnus arm, and has only one arm tangent as drawn from the Sun (located in galactic Quadrant IV). Angles a and b in this plot are for mathematical uses here in equations; they should not be used to describe physical galactic latitude. Adapted and updated from fig.1 in Vallée (2015).

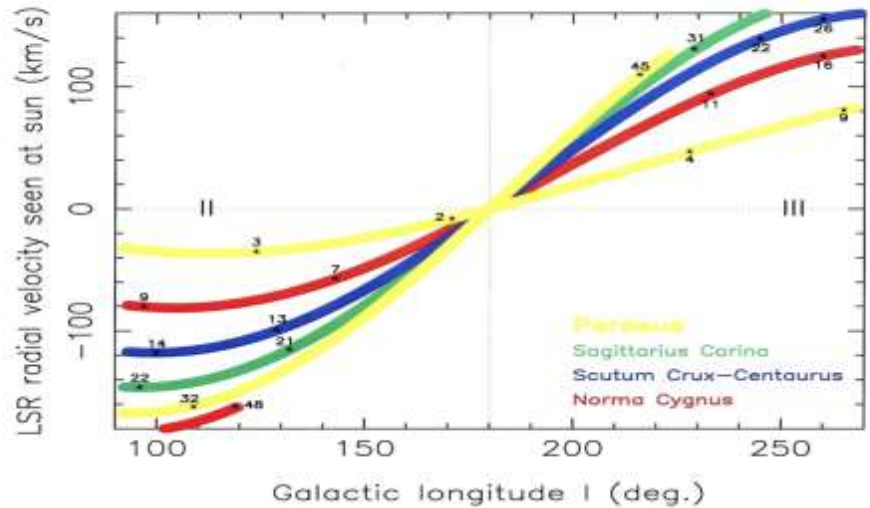

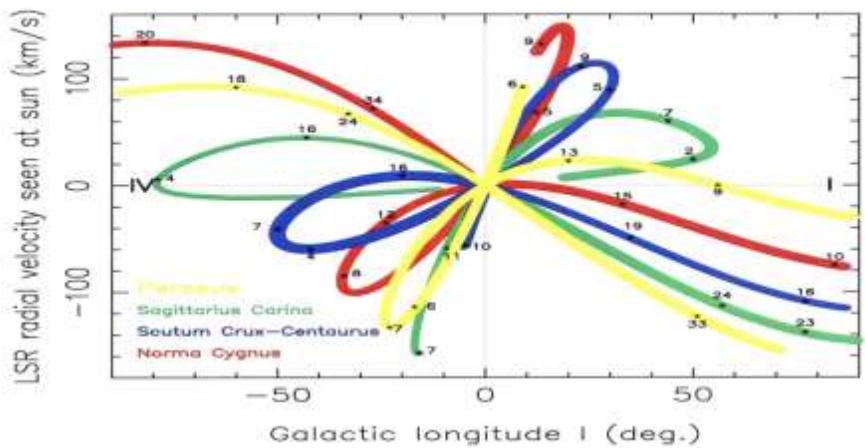

**Figure 3.** Sketch of the radial velocity of each spiral arm, seen at different galactic longitudes (using a flat circular velocity of 220 km/s). Each logarithmic spiral arm is colored as in Figure 1. Pitch angle of 13.1° inward, arm starting at 2.2 kpc. On each spiral arm, numbers indicate the distance from the Sun, with the GC-Sun distance being 8.0 kpc.
Adapted and updated from figure 3 in Vallée (2008a).
(a) Toward the Galactic Anticentre. Galactic quadrants III (left) and II (right) are shown.
(b) Toward the Galactic Center. Galactic quadrants IV (left) and I (right) are shown. The Galactic Center is at (0,0). The negative longitude values shown here represent the amount to be subtracted from 360° to yield their positive values.

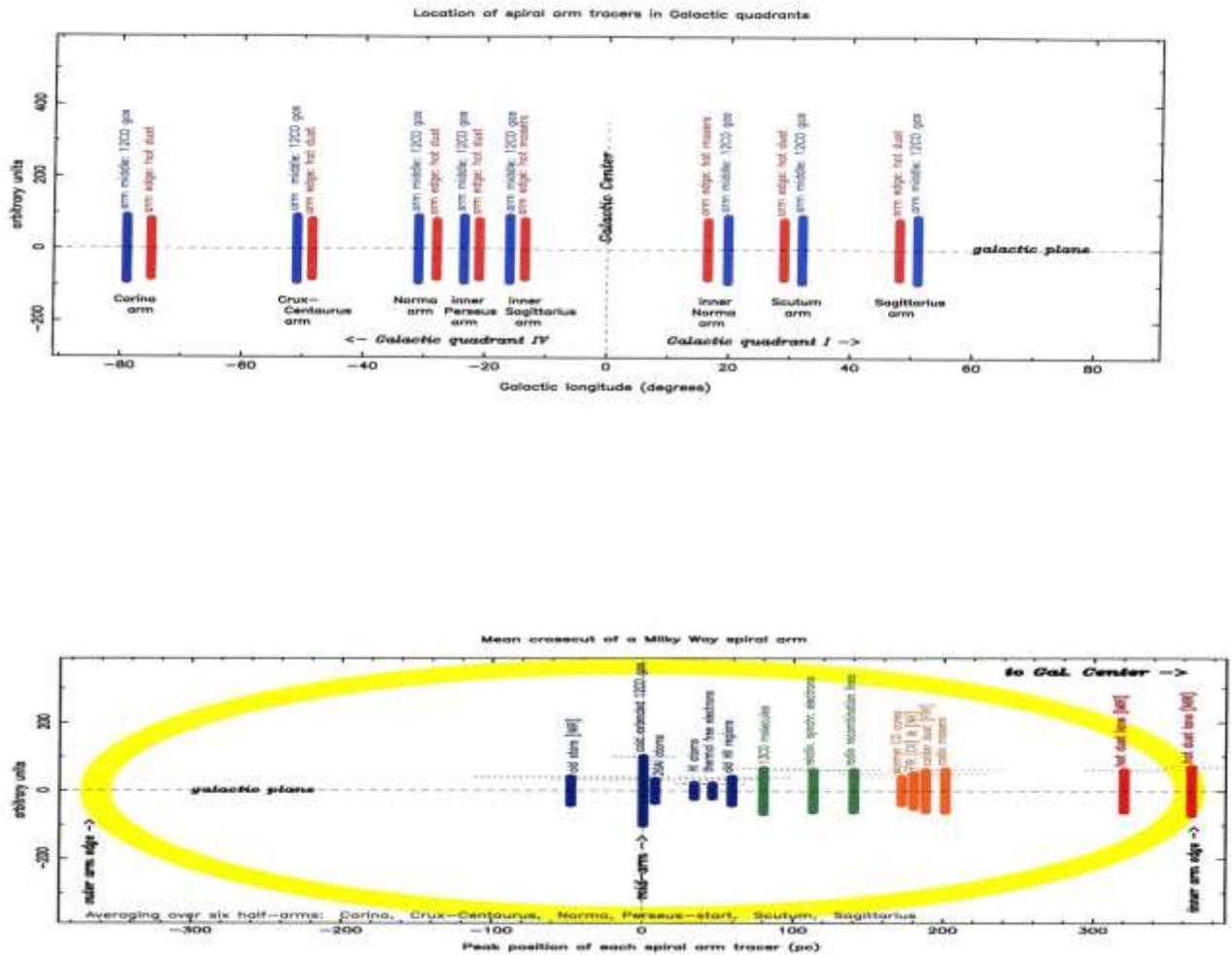

**Figure 4.** Display of spiral arm tracers.
Adapted and updated from figs 1 and 2 in Vallée (2016b).

**a)** Display of the mean longitudes of hot dust tracers and of CO tracers, as a function of galactic longitude. A Sun-GC distance of 8.0 kpc is employed.

**b)** Crosscut of a typical spiral arm (yellow ellipse), from the outer edge (at left) to the mid-arm (0), to the inner edge (at right). The separation S of each tracer from the origin (mid-arm) is shown (data from Table 2, assuming the GC-Sun distance of 8.0 kpc). Vertical units are arbitrary. In a spiral arm, the inner edge is closest to the Galactic Center. This sketch incorporates 6 spiral arms: Scutum, Sagittarius, Carina, Crux-Centaurus, Norma, Perseus origin. New color coding starts from the mid-arm (blue) for cold $^{12}$CO, old stars in near-infrared, visible HII regions, all within the central zone within their errors; this is followed with offsets (green) for relativistic synchrotron electron, $^{13}$CO; more offsets (orange) follow for FIR [CII] & [NII] lines, cold dust seen in FIR wavelengths, masers; this ends with the largest offsets (red) for the hot dust lanes seen in MIR and NIR wavelengths at about 340 pc (or 1100 light-years).

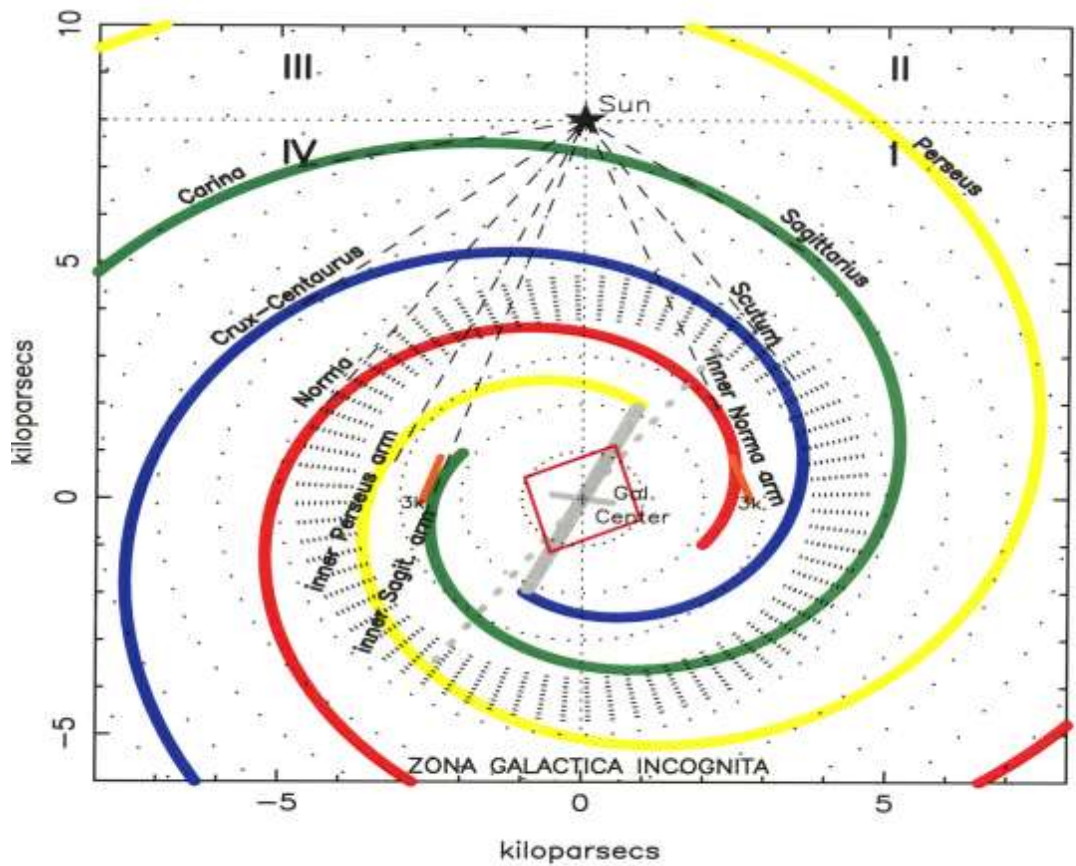

Figure 5. Sketch of the starts of the 4 spiral arms. Arm tangents are shown as dashed lines from the Sun. Same color coding as in Figures 1 and 3. The model-predicted arm tangents for the start of Sagittarius arm (green) and for the start of Norma arm (red) are shown, as well as the observed tangents for the start of Perseus arm (yellow) and for the start of Scutum arm (blue). Galactic quadrants are shown (I to IV). The Galactic Center is at (0,0), and the Sun at (0,8). The galactic bulge is shown by a diamond, oriented along the broad gray bar at $25^o$ from the line joining the Sun to the Galactic Center. Two other putative bars are shown: the thin long bar (dashed) and the short nuclear bar. Also shown are the putative molecular ring (radial dots) around 4 kpc, and the mean location of tangents to the so-called ' 3-kpc-arm' features (orange). The arm pitch angle is $-13.1^o$. Not much is observed behind the Galactic Center (*Zona Galactica Incognita*). Adapted and updated from fig. 5 in Vallée (2016a).

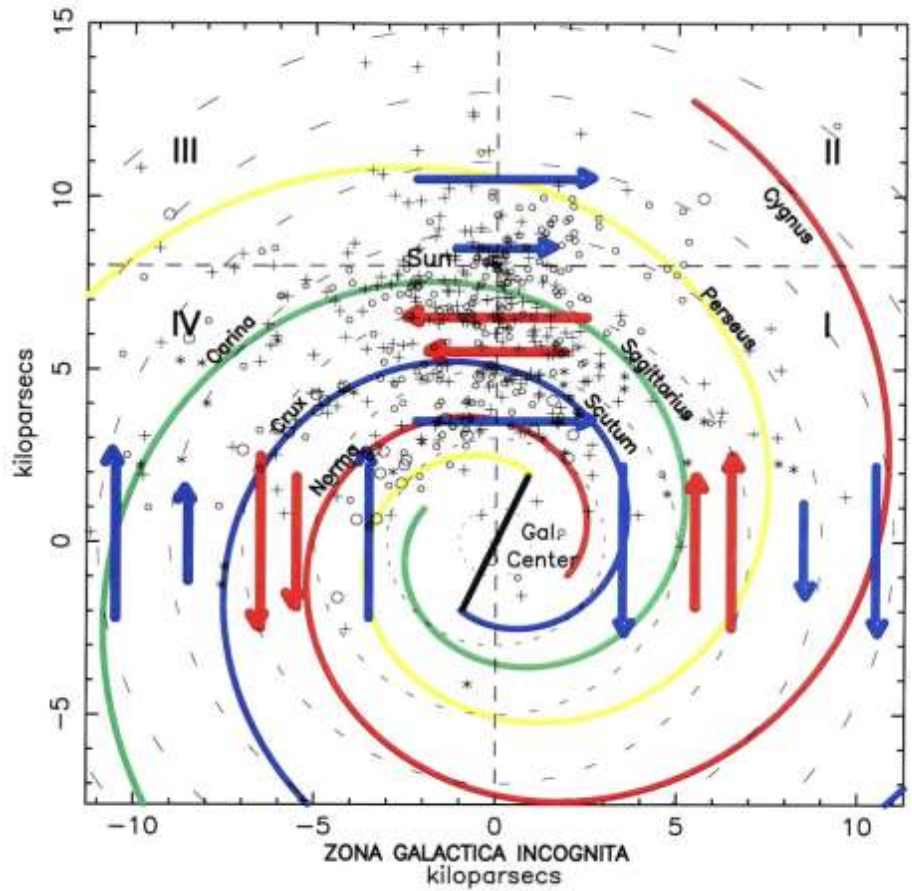

Figure 6. Sketch of the magnetic field vectors. Pulsar Rotation Measures (RM) are shown by + sign or * sign (small or large positive RM) or by small and large circles (small or large negative RM). Magnetic field vectors are shown by arrows (blue going clockwise; red going counterclockwise). The global disk magnetic field is essentially clockwise (blue), except for a small zone where the magnetic field has reversed (red). Large dashed circles turn around the Galactic Center (circular orbits). Model parameters as follows: Sun-GC distance now at 8.0 kpc; spiral arm pitch now at -13.1°. Magnetic data below the Galactic Center (*Zona Galactica Incognita*) are unavailable; those at left and right near the x-axis may be less precise (they have a small pitch angle). Adapted and updated from fig. 5 in Vallée (2008b).